# BOTDA Fiber Sensor System Based on FPGA Accelerated Support Vector Regression

Huan Wu[†], Hongda Wang[†], Chiu-Sing Choy, *Senior Member, IEEE*, Chester Shu, *Senior Member, IEEE*, and Chao Lu, *Fellow, OSA*

[†] These authors contribute equally to this work

*Abstract*—Brillouin optical time domain analyzer (BOTDA) fiber sensors have shown strong capability in static long haul distributed temperature/strain sensing. However, in applications such as structural health monitoring and leakage detection, real-time measurement is quite necessary. The measurement time of temperature/strain in a BOTDA system includes data acquisition time and post-processing time. In this work, we propose to use hardware accelerated support vector regression (SVR) for the post-processing of the collected BOTDA data. Ideal Lorentzian curves under different temperatures with different linewidths are used to train the SVR model to determine the linear SVR decision function. The performance of SVR is evaluated under different signal-to-noise ratios (SNRs) experimentally. After the model coefficients are determined, algorithm-specific hardware accelerators based on field programmable gate arrays (FPGAs) are used to realize SVR decision function. During the implementation, hardware optimization techniques based on loop dependence analysis and batch processing are proposed to reduce the execution latency. Our FPGA implementations can achieve up to 42x speedup compared with software implementation on an i7-5960x computer. The post-processing time for 96,100 BGSs along 38.44-km FUT is only 0.46 seconds with FPGA board ZCU104, making the post-processing time no longer a limiting factor for dynamic sensing. Moreover, the energy efficiency of our FPGA implementation can reach up to 226.1x higher than software implementation based on CPU.

## I. INTRODUCTION

SINCE the invention of Brillouin optical time domain analyzer (BOTDA) in 1990 [1], it has attracted both academic and industrial interests [2-4]. BOTDA sensors rely on stimulated Brillouin scattering (SBS) of two counter-propagating light waves, a continuous-wave (CW) signal and a pulsed pump. The frequency offset between the pump and probe is scanned around the Brillouin frequency shift (BFS) of the fiber to reconstruct the Brillouin gain spectrum (BGS). Since the change of BFS has a linear relationship with the change of temperature and strain on the fiber, an important operation in a BOTDA system is to find the BFS from the measured BGS to determine temperature or strain information along the fiber under test (FUT). In an ideal BGS, BFS is the shift in peak gain frequency. However, acquired BGSs are always contaminated by noises. Therefore, post-processing algorithms are needed to accurately determine BFS from the measured BGSs. The conventional wisdom to predict the BFS information from the BGS is Levenberg-Marquardt algorithm (LMA) curve fitting [5-6]. However, its complexity is often a limiting factor in the sensing speed of a BOTDA system especially for long sensing distance.

In recent years, the performances of BOTDA are improved significantly due to the rapid developments of the technology. The sensing distance of BOTDA can achieve hundreds of kilometers [7], and the spatial resolution can be reduced to millimeter level [8]. Longer sensing distance brings larger amount of sensing data and finer resolution requires higher sampling rate and smaller frequency scanning step which result in denser sensing points. The sensing data volume keeps increasing, which adds the computational load for post-processing. In real scenario, to extract temperature/strain information from the measured BGSs with low latency is quite necessary. However, traditional LMA curve fitting technique is time consuming due to its iterative nature. Several works have mentioned the challenges of post-processing in real applications. In [9], a non-curve fitting technique called cross-correlation method (XCM) was proposed based on calculation of cross-correlation between an ideal Lorentzian curve and the measured BGS to determine BFS. In [10], a modified version of XCM was implemented on FPGA to speed up the processing time. Artificial neural network (ANN) is also proposed for BOTDA system to improve the sensing accuracy and processing speed [11]. However, the training of ANN is difficult due to numerous hyperparameters. Recently, we reported a machine learning method called support vector machine (SVM) to extract temperature information from measured BGSs with simple training strategy and fast processing speed [12-14].

SVM was first introduced by Vapnik in 1963 [15]. It is a powerful and versatile machine learning algorithm, capable of performing linear or nonlinear classification and regression. SVM for classification is called support vector classification (SVC) and for regression is called support vector regression

This work was supported by CUHK Group Research Scheme, Research Grants Council of Hong Kong (RGC) project: RGC GRF CUHK 14204918 and PolyU 152658/16.

Huan Wu and Chao Lu are with Department of Electronic and Information Engineering, The Hong Kong Polytechnic University, Kowloon, Hong Kong. (email: nuaawuhuan@gmail.com)

Hongda Wang, Chiu-Sing Choy and Chester Shu are with the Department of Electronic Engineering, The Chinese University of Hong Kong, Shatin, N.T., Hong Kong. (email: 1155039965@link.cuhk.edu.hk).

(SVR). SVC solves binary classification problems by formulating them as convex optimization problems, and the optimization aims to find the maximum margin separating the hyperplane while correctly classifying as many training points as possible. In [12-14], we treat extracting temperature information from BGSs as a supervised classification problem, the BGSs serving as feature vectors are classified into different temperature classes by the SVC model. SVM can also be used as regression method. As opposed to SVC which can only output a discrete value, SVR returns a continuous-valued output. Since temperature and strain on the fiber are continuous values, SVR is more suitable for BOTDA data. In this work, we use SVR to extract continuous temperature information from BGS and further improve the post-processing speed by introducing a hardware accelerator based on field-programmable gate array (FPGA). The main contributions of this work are as follows:

1) A new temperature prediction method based on SVR is proposed. Unlike SVC, which can only output discrete temperatures, SVR can predict continuous temperature information from measured BGSs acquired from a BOTDA system. The experimental results prove that SVR can achieve comparable performance with SVC under different SNRs. However, SVR is more suitable for hardware implementations.

2) Hardware implementations of SVR decision function are realized on two FPGA boards. Optimizations to linear SVR decision function through loop analysis and batch processing are proposed to take advantages of high flexibility and scalability of modern FPGA devices. These optimization methods transform the decision function into matrix-matrix multiplication and matrix-vector multiplication and parallelize these operations by tiling the large matrix into smaller ones.

3) Post-processing time for 96,100 BGSs along 38.44-km FUT can be completed in 0.46 seconds with Xilinx ZCU104 using the proposed hardware optimization techniques. It achieves 42x speedup compared with the software implementation running on an i7-5960x computer. Meanwhile, the 26.5W power consumption of ZCU104 is also much lower than the conventional CPU, making the energy efficiency of our FPGA implementation 221.6x higher than software implementation based on LIBSVM [16].

The paper is organized as follows. Section II describes the principle of SVR and its training process for temperature extraction in a BOTDA system. Section III introduces the experimental setup of BOTDA and evaluates the performance of SVR under different SNRs experimentally. FPGA optimizations and implementations of linear SVR decision function are given in Section IV. Section V concludes this work.

## II. Principle of SVR and Training Process for Temperature Extraction in a BOTDA System

Suppose we have training data $\{(x_1, y_1), \ldots, (x_l, y_l)\}$, where $x_i \in R^n$ is training sample and $y_i \in R$ is label. In linear case, we construct a linear decision function to fit the training data:
$$f(x) = \langle w, x \rangle + b \quad (2.1)$$
where $\langle \cdot, \cdot \rangle$ denotes the dot product, $w$ is the norm vector of the linear function and $b$ is intercept. Traditional linear least-square error regression derives a decision function by minimizing the deviation between predicted value $f(x_i)$ and given value $y_i$ for all training data. Unlike linear least-square error fitting, SVR allows a tolerance degree to errors not greater than $\varepsilon$ as shown in Fig. 1(a). Only the data points outside the shaded region contribute to the error and the deviations are penalized in a linear fashion as shown in Fig. 1(b). The goal of SVR is to find a function that fits current training data with a deviation no larger than $\varepsilon$, and at the same time as flat as possible. One way to ensure this is to minimize the norm, i.e., $\|w\|^2 = \langle w, w \rangle$. We can write this problem as a convex optimization problem as follows:

$$\text{minimize: } \frac{1}{2}\|w\|^2$$
$$\text{subject to } \begin{cases} y_i - \langle w, x \rangle - b \leq \varepsilon \\ \langle w, x \rangle + b - y_i \leq \varepsilon \end{cases} \quad (2.2)$$

The above convex optimization problem is feasible in cases where $f(x)$ actually exists and all pairs $(x_i, y_i)$ are within $\varepsilon$ precision. However, in most cases, not all $(x_i, y_i)$ are within $\varepsilon$ precision, then we can introduce slack variables $\xi_i, \xi_i^*$ to deal with this problem. Hence, we get the following formulation:

$$\text{minimize: } \frac{1}{2}\|w\|^2 + C \sum_{i=1}^{l}(\xi_i + \xi_i^*)$$
$$\text{subject to } \begin{cases} y_i - \langle w, x \rangle - b \leq \varepsilon + \xi_i \\ \langle w, x \rangle + b - y_i \leq \varepsilon + \xi_i^* \end{cases} \quad (2.3)$$

where $\xi_i, \xi_i^* \geq 0$, the constant $C > 0$ determines the trade-off between the flatness of $f(x)$ and the amount up to which deviations larger than $\varepsilon$ are tolerated. Equation (2.3) is known as the primal problem of SVR algorithm and it can be transformed to dual problem and solved by quadratic programming [17]. The solution is as follows:

$$\mathbf{w} = \sum_{i=1}^{l}(\alpha_i - \alpha_i^*)(x_i) \quad (2.4)$$
$$f(x) = \sum_{i=1}^{l}(\alpha_i - \alpha_i^*)\langle x_i, x \rangle + b \quad (2.5)$$

where $\alpha_i$ and $\alpha_i^*$ are the dual variables, $\langle x_i, x \rangle$ represents the inner product between training sample $x_i$ and test sample $x$. From Equation (2.5), we can see that once the model parameters are identified, SVR only depends on $x_i$ with corresponding $(\alpha_i - \alpha_i^*)$ which are non-zero, these $x_i$ are called support vectors and they are subsets of training data.

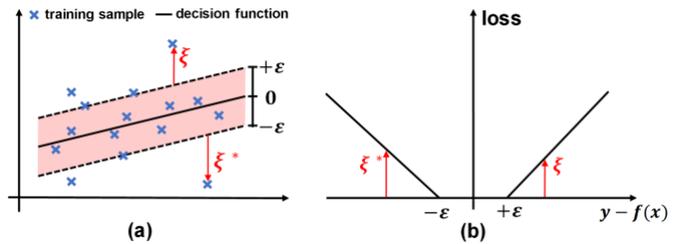

Fig. 1. (a) one-dimensional linear SVR, (b) linear loss function.

In our case, to process measured BGSs collected from a BOTDA system, a high dimensional linear SVR is used, normalized gain value at every frequency on the BGS forms feature vector $x_i$, and corresponding temperature of the BGS is label $y_i$. The use of SVR includes two phases, the training phase and testing phase as shown in Fig. 2. During the training phase, the simulated ideal BGSs together with the corresponding temperature labels serving as the training samples are used to get linear decision function for temperature prediction. We design the simulated ideal BGSs by using ideal Lorentzian curve as the gain profile for the training of SVR:

$$g(\nu) = \frac{g_B}{1+\left[\frac{(\nu-\nu_B)}{\Delta\nu_B/2}\right]^2} \quad (2.6)$$

where $g_B$, $\nu_B$ and $\Delta\nu_B$ are the peak gain, BFS and bandwidth of the BGS. Peak gain is set as 1, BFSs of the ideal BGSs from a temperature range of 0℃ to 70℃ with 0.5℃ step are determined using the temperature coefficient of the fiber under test (FUT). The linewidth of ideal BGSs vary from 30MHz to 100MHz at a step of 2MHz to accommodate BGS linewidth variation in experiment. Finally, we have $141 \times 36$ ideal BGSs to train the SVR. The frequency range of $\nu$ is from 10.78GHz to 11.0GHz with 1MHz step, therefore, we have 220 frequencies. After training, we get 1,136 support vectors in the SVR model. In the testing phase, the fixed model predicts a continuous temperature value for each normalized measured BGS collected from a BOTDA system.

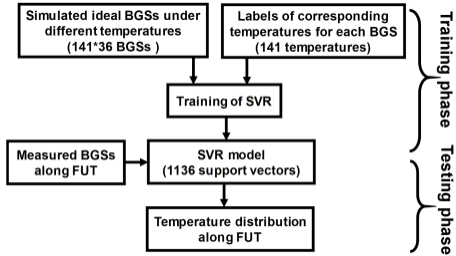

Fig. 2. Training and testing phase of SVR.

## III. BOTDA Setup and Experimental Results

### A. BOTDA Experimental Setup

The experimental setup of the BOTDA system is shown in Fig. 3. The output of a tunable laser source is set around 1550nm and is split into two branches using a coupler. The CW light in the upper branch is modulated by a Mach-Zehnder modulator (MZM) driven by a pulse pattern generator (PPG) to generate optical pump pulses. The bias controller after MZM is to stabilize the applied voltage. The pump is then amplified by an erbium-doped fiber amplifier (EDFA) and passes through a polarization scrambler (PS) to eliminate polarization dependent noise. In the lower branch, another high extinction ratio MZM is driven by a radio frequency (RF) generator. The bias controller is biased at Null point to generate a carrier suppressed double-sideband probe signal. An optical attenuator (ATT) is used to control the probe power followed by an isolator to block the signal from the pump branch. The probe signal is detected by a photodetector (PD) after the lower-frequency probe sideband is selected by using a fiber Bragg grating (FBG) filter. Local BGSs are reconstructed with RF scanned around the BFS of FUT. Ensemble average is commonly used in BOTDA to increase SNR at the expense of longer data acquisition time.

Temperature/strain measurement time of the BOTDA system includes the data acquisition time $T_{acq}$ and post-processing time $T_{pp}$, and can be expressed as follows:

$$T = T_{acq} + T_{pp} = (T_c \cdot N_{avg} + T_s)N_{freq} + T_{pp} \quad (3.1)$$

where $T_c = 2nL/c$ is time of flight, $L$ is the length of FUT, $n$ is the refractive index of the fiber and $c$ is light speed in vacuum. $N_{avg}$ is the number of averages, $T_s$ is the frequency switching time of RF which is around hundreds of milliseconds and $N_{freq}$ is the number of scanned frequencies.

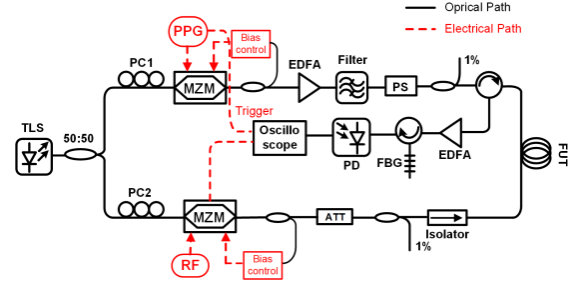

Fig. 3. BOTDA experimental setup. TLS: tunable laser source, PC: polarization controller, PPG: pulse pattern generator, RF: radio frequency, PS: polarization scrambler, MZM: Mach-Zehnder modulator, ATT: attenuator, FUT: fiber under test, FBG: fiber-Bragg grating, PD: photodetector.

### B. Experimental Results

To evaluate the performance of SVR, we use the BOTDA setup in Fig. 3 to measure the BGS distribution along 38.44-km FUT. The last 400-m section of FUT is free from strain and put in a temperature oven heated to 50℃ as shown in Fig. 4(a). The sampling rate is 250MSample/s, corresponding to 96,100 sampling points for 38.44-km FUT. Fig. 4(b) shows the BGSs distribution measured with 20ns pump pulse, 1024 times averaging, and the sweeping frequency is from 10.78GHz to 11.0GHz with 1MHz frequency step.

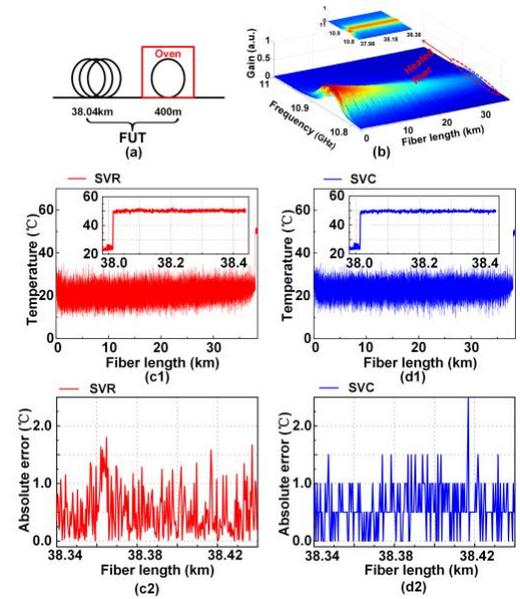

Fig. 4. (a) 38.44-km FUT with last 400m heated to 50℃. (b) Measured BGS distribution along FUT. Temperature distribution along FUT determined by (c1) SVR and (d1) SVC, insets: zoom-in view at the heated section. Absolute temperature error along 100-m FUT by (c2) SVR and (d2) SVC.

Next, the measured BGSs are processed by SVR. For comparison, we also process the BGSs by SVC. The extracted temperature distributions by SVR and SVC are shown in Fig. 4(c1) and (d1), respectively. Both the training data and testing data are same for SVR and SVC. The insets in Fig. 4(b), (c1) and (d1) depict the zoom-in view at the heated section. We can see that the temperature information along FUT has been successfully extracted by both SVR and SVC. SVR can achieve comparable performance as SVC, the temperature uncertainty at the last 400-m FUT are 0.608℃ for SVR and 0.549℃ for SVC, respectively. Fig. 4(c2) and (d2) show the absolute error of the predicted temperature by SVR and SVC from 38.34 km to 38.44 km, we can clearly see that the predicted temperature from SVR are continuous values while that from SVC are discrete values, and they exhibit similar error fluctuation range and prediction capability.

SVC, respectively. Fig. 4(c2) and (d2) show the absolute error of the predicted temperature by SVR and SVC from 38.34 km to 38.44 km, we can clearly see that the predicted temperature from SVR are continuous values while that from SVC are discrete values, and they exhibit similar error fluctuation range and prediction capability.

Then we investigate the tolerance of SVR to different level of SNRs, the pump pulse is fixed at 20ns and frequency scanning step is 1MHz. SNR is defined as the ratio between the mean amplitude of Brillouin peak and its standard deviation [18], which is proportional to the amplitude instead of power. We collect the BGSs from 4.5dB to 12dB by using 32 to 1024 times of averaging. According to Equation (3.1), theoretical measurement time varies from 2.7 seconds to 88 seconds when averaging time increases from 32 to 1024. Fig. 5 shows the temperature uncertainty predicted by SVR and SVC under different SNRs. We can see that lower uncertainty can be achieved with higher SNR for both SVR and SVC at the expense of longer data acquisition time. While at a same SNR, SVR and SVC have comparable performance.

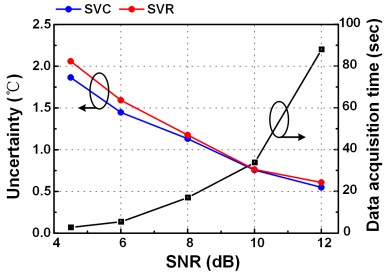

Fig. 5. Temperature uncertainty around the last 400-m section by using SVC and SVR for temperature extraction.

## IV. FPGA Optimizations and Implementations of SVR

FPGA can produce very strong computation capability through parallelizing the algorithm in an efficient manner. Moreover, compared with other hardware accelerators like application-specific integrated circuit (ASIC), FPGA also have the advantages of reconfigurability and fast deployment time especially with the help of high level synthesis (HLS) [19]. However, not all the algorithms can achieve real-time acceleration because of the inadaptability to fixed hardware structures. In [11], the authors use ANN to predict the temperature information and the performance improvement over LMA curve fitting technique is remarkable. However, from the hardware perspective, the sigmoid nonlinear activation function in each neuron is very expensive to realize, thus ANN is not very suitable for efficient hardware implementation. As shown in Section III, both SVC and SVR can be used to extract temperature information from BOTDA with excellent performance, however, n-class SVC is built upon $n(n-1)/2$ binary classifiers and each classifier has unique number of support vectors, the irregular computation pattern doesn't fit a fixed hardware structure. While SVR predicts the result by regular matrix-vector multiplication and inner-product, which is very suitable to be parallelized and pipelined from the hardware perspective. With a dedicated FPGA accelerator, the processing speed of linear SVR can be significantly improved.

In this section, a hardware architecture for the linear SVR decision function is presented. In the following subsections, part A introduces the direct implementation of linear SVR decision function and discusses its drawbacks. In part B, optimizations to direct implementation by loop analysis are proposed to reduce the latency. In part C, batch processing method is proposed to further speed up the running time. In part D, 96,100 measured BGSs from 38.44-km FUT are processed by two FPGA boards, experimental results and comparison with software implementation are described. In part E, we give an in-depth theoretical analysis and discussion for FPGA acceleration with the proposed optimization techniques.

### A. Direct Implementation of Linear SVR Decision Function

If we simplify $(\alpha_i - \alpha_i^*)$ in decision function Equation (2.5) as $\beta_i$ and expand the inner product to a sum-of-product term, then we can have the reformulated decision function as follows:

$$f(x) = \sum_{i=1}^{N_s} \beta_i \sum_{j=1}^{M} SV_{ij} x_j + b \quad (4.1)$$

where $SV$ represents support vectors obtained from the training process, $N_s$ is the number of support vectors and is 1136 as given in Section II, $M$ is the dimension of input feature vector and is equal to 220. The data path of Equation (4.1) can be illustrated in Fig. 6 and the corresponding pseudocode is shown in Algorithm 1. In Fig. 6, multiply-accumulate (MAC) 1 corresponds to the inner summation of Equation (4.1) and is denoted as partial sum, while MAC 2 corresponds to the outer summation and is denoted as final sum. The total MAC operations in MAC 1 and MAC 2 are $(N_s M + N_s) * 2 * N_{BGS}$, where $N_{BGS}$ is the number of BGSs. To process 96,100 measured BGSs from 38.44-km FUT, about $2.4 \times 10^{10}$ multiplications and $2.4 \times 10^{10}$ summations are needed, resulting in a heavy computation burden for real-time processing.

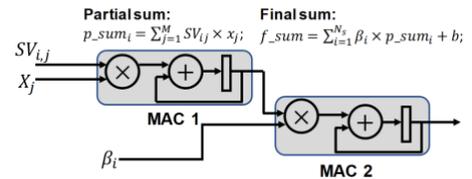

Fig. 6. Data path of linear SVR.

---
**Algorithm 1:** Original linear SVR without optimization
---
**Input**: feature vector $x[M]$
**Require**: support vectors $SV[N_s][M]$, support vector corresponding multipliers $\beta[N_s]$, bias
**Output**: regression result $f(x)$
**Initialize**: $p\_sum[N_s] \leftarrow 0, f\_sum \leftarrow bias$
**L1: for** $i=0$ to $N_s - 1$ **do**
    **L2: for** $j=0$ to $M - 1$ **do**
        $square \leftarrow SV[i][j] * x[j]$;
        $p\_sum[i] \leftarrow p\_sum[i] + square$;
    **end for**
    $temp[i] \leftarrow \beta[i] * p\_sum[i]$;
    $f\_sum \leftarrow f\_sum + temp[i]$;
**end for**
$f(x) \leftarrow f\_sum$;

---

In hardware design, parallel and pipeline are two common techniques to reduce the latency. However, the loop-carried dependence in the inner loop L2 causes long pipeline initiation interval and inefficient hardware utilization efficiency. Moreover, due to the existed dependence, parallelism of this direct implementation cannot be achieved without restructuring the code, thus the total latency is heavily restricted. To accelerate the decision function and enable real-time processing, optimizations must be performed to overcome the limitations.

*B. Loop Dependence Analysis and Optimizations*

To remove the loop-carried dependence and parallelize the partial sum computation, firstly, we need to perform loop dependence analysis [20]. In Algorithm 1, the statements inside L2 exhibit inter-dependence with respect to the iterator $j$, but show no inter-dependence on iterator $i$. Thus, we seek to change the execution order of L1 and L2 to remove the inter-dependence. However, the nested loop is imperfect (perfect nested loops mean the statements only exist inside the innermost loop), we need to take a two-step optimization.

- **Loop distribution:** We find that the statements inside L2 do not depend on the statements between L1 and L2, this means we can safely break loop L1 and distribute the statements between L1 and L2 outside. After loop distribution, a new loop L3 is formed which is only responsible for the accumulation of final sum, while L1 and L2 become a perfect nested loop and calculates the partial sum.
- **Loop interchange:** In the perfect nested loop L1 and L2, loop-carried dependence prevents efficient pipeline strategy to be applied because of the long execution latency of the accumulator. The pipeline initiation interval is restricted by the propagation delay of the adder, which is normally larger than one clock cycle for floating point numbers. When working in higher frequency, the propagation delay could further consume more clock cycles, resulting in longer pipeline initiation interval. In Algorithm 1, no inter-dependence is observed between the statements inside L2 and the iterator $i$, therefore, we can interchange L1 and L2 to remove the dependence and make the nested loop executed consecutively in each clock cycle. After loop interchange, the partial sum is read and write simultaneously with no conflict on the access addresses, which indicates that the partial sum should be mapped to the dual port RAM on FPGA.

---

**Algorithm 2:** Optimized linear SVR with loop distribution and loop interchange

---

**Input**: feature vector $x[M]$
**Require**: support vectors $SV[M][N_s]$, support vector corresponding multipliers $\beta[N_s]$, bias
**Output**: regression result $f(x)$
**Initialize**: $p\_sum[N_s] \leftarrow 0$, $f\_sum \leftarrow bias$
**L1: for** $i=0$ to $M-1$ **do**
  **L2: for** $j=0$ to $N_s-1$ **do**     ◁ loop unroll
    $square \leftarrow SV[i][j] * x[i]$;
    $p\_sum[j] \leftarrow p\_sum[j] + square$;
  **end for**
**end for**
**L3: for** $i=0$ to $N_s-1$ **do**     ◁ loop unroll
  $temp[i] \leftarrow \beta[i] * p\_sum[i]$;
  $f\_sum \leftarrow f\_sum + temp[i]$;
**end for**
$f(x) \leftarrow f\_sum$;

---

The pseudocode after loop distribution and loop interchange is in shown Algorithm 2. Since the execution order of L1 and L2 is changed, the support vector matrix also needs to be transposed accordingly. The total execution latency in clock cycles can be expressed as follows:

$$\text{Latency} = N_s M + N_s T_a \quad (4.2)$$

where $T_a$ is the propagation delay of the adder.

Parallelization is another advantage after eliminating loop-carried dependence by loop distribution and interchange. In Algorithm 2 we know that $p\_sum[j]$ and $p\_sum[j+1]$ are calculated independently, thus we can unroll the loop L2 directly to increase the parallelism without changing the code structure. After unrolling, massive parallelized MAC units can be mapped to DSP slices on FPGA easily. Meanwhile, same level of parallelism can also be applied to L3 to shorten the latency. Assume we unroll L2 and L3 with a factor of $f$ and the delay of an adder is $T_a$, the total latency can be calculated as follows:

$$\text{Latency} = \underbrace{\frac{N_s M}{f}}_{\text{Partial sum}} + \underbrace{fT_a + \frac{N_s}{f} + L_{tree}(f)}_{\text{Final sum}} \quad (4.3)$$

$$L_{tree}(f) \approx \begin{cases} T_a \left\lceil \log_2 \frac{N_s}{f} \right\rceil, & f > \frac{N_s}{2f} \\ T_a \left\lceil \log_2 \frac{N_s}{f} \right\rceil + 2 \left\lceil \frac{N_s}{2f^2} \right\rceil - 2, & 1 < f < \frac{N_s}{2f} \end{cases}$$

where $L_{tree}(f)$ is the latency of the adder tree inside L3 after unrolling. $L_{tree}(f)$ has different expressions with small and large unroll factors, but in both cases it has little effect on total latency, therefore it can be dropped safely in later analysis. Note that the latency for $f=1$ is calculated separately as Equation (4.2). To study the effect of parallelization, we apply different unroll factors on Algorithm 2. The target platform is Xilinx ZCU104 and the working frequency is set to 200 MHz. All the input signals and intermediate values use single-precision floating point numbers. The execution latency and speedup factor are collected from Vivado HLS synthesis report, shown in Fig. 7(a). We can see that the latency for one regression decreases fast as the unroll factor increases, and the speedup almost scales linearly when the unroll factors are relatively small ($\leq 36$). But if we further increase the unroll factor, the linear scaling does not hold and the acceleration effect is weakened. When the unroll factor increases to 142, the real speedup is about 92x. This can be explained by the following equation:

$$\text{Latency} \approx \begin{cases} \frac{N_s M}{f}, & \text{for small } f \\ \frac{N_s M}{f} + fT_a + \frac{N_s}{f}, & \text{for large } f \end{cases} \quad (4.4)$$

For small unroll factor, the latency for final sum calculation is negligible compared with partial sum, therefore the total latency is approximately inversely proportional to $f$. For large unroll factor, the latency of the adder chain within L3 is comparable to that of the partial sum, so the linear scaling does not hold anymore. The hardware consumption is shown in Fig. 7(b), we can see the DSP consumption scales linearly with the unroll factor, while the block-RAM (BRAM) consumption doesn't change much because the support vectors dominate most of the BRAM usage. The look-up table (LUT) and flip-flop (FF) consumptions are also proportional to the unroll factor. The results prove that area-performance trade-off can be easily achieved with the proposed optimization method.
5

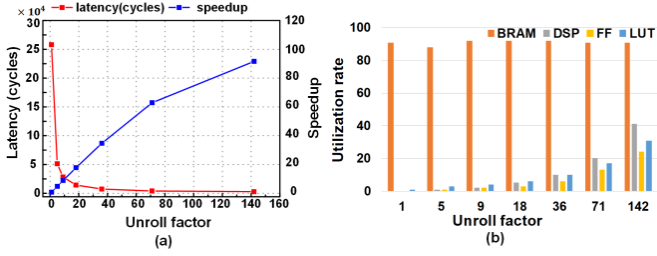

Fig. 7. (a) Speedup and latency versus unroll factor, (b) hardware utilization rate on ZCU104 versus unroll factor.

## C. Batch Processing Method

From the experimental results in Fig. 7(a), we know that the latency can be greatly reduced with the proposed two-step optimization method and loop unroll, thus a notable speedup can be achieved. However, the linear scaling relationship is not valid for large unroll factor. If we want to achieve high parallelism with a large unroll factor, the latency of $fT_a$ of the long adder chain becomes prominent, since it is proportional to the unroll factor $f$. Under this circumstance, very long pipeline stages of the adder chain in L3 will cause the MAC units under-utilized.

To further improve the hardware utilization efficiency of L3 with large unroll factor, we propose a batch processing method to process a batch of input vectors at a time. With batch processing, the nested loop L1 and L2 in Algorithm 2 become a three-level nested loop L1, L2 and L3, while the original L3 loop turns into a nested loop L4 and L5. The pseudocode for batch processing is shown in Algorithm 3. The total latency of Algorithm 3 can be calculated as follows:

$$\text{Latency} \approx \underbrace{\frac{BN_sM}{f}}_{\text{Partial sum}} + \underbrace{fT_a + \frac{BN_s}{f}}_{\text{Final sum}} \quad (4.5)$$

where B is the batch size. If we divide the total latency by B, the average latency of the adder chain inside L3 is now shared by B inputs:

$$\text{Average Latency} \approx \frac{N_sM}{f} + \frac{fT_a}{B} + \frac{N_s}{f} \quad (4.6)$$

When $B$ increases, the average latency of the adder chain will decrease and finally we can have the approximate average latency as follows when $B$ is large enough:

$$\text{Average Latency} \approx \frac{N_s(M+1)}{f} \quad (4.7)$$

In Equation (4.7), we can see the latency is only dependent on the unroll factor $f$, which exhibits an inversely proportional relationship and the linear scaling of speedup holds.

The hardware structure for calculating the three-level nested loop in Algorithm 3 is shown in Fig. 8. To enable multiple access to the support vector matrix, array partition is performed to increase the memory bandwidth and the partition factor is equal to the unroll factor. Moreover, the partitioned partial sum matrix is mapped to the dual port RAM to enable simultaneous read and write operations. In every clock cycle, $f$ support vectors and one element from input vectors are read to the parallel MAC array, the accumulation results are written to the dual port RAM concurrently. It takes totally $BN_sM/f$ cycles to finish updating the partial sum matrix. After this, the partial sum matrix will be used to calculate the final sum.

**Algorithm 3:** Optimized linear SVR with loop distribution, loop interchange and batch processing

**Input**: multiple feature vectors $x[B][M]$
**Require**: support vectors $SV[M][N_s]$, support vector corresponding multipliers $\beta[N_s]$, bias
**Output**: classification results $f(x[B])$
**Initialize**: $p\_sum[N_s] \leftarrow 0, f\_sum[B] \leftarrow bias$
**L1: for** $k$=0 to $B-1$ **do**
    **L2: for** $i$=0 to $M-1$ **do**
        **L3: for** $j$=0 to $N_s-1$ **do** ◁ loop unroll
            $square[k] \leftarrow SV[i][j] * x[k][i]$;
            $p\_sum[k][j] \leftarrow p\_sum[k][j] + square[k]$;
        **end for**
    **end for**
**end for**
**L4: for** $k$=0 to $B-1$ **do**
    **L5: for** $i$=0 to $N_s-1$ **do** ◁ loop unroll
        $temp[k][i] \leftarrow \beta[i] * p\_sum[k][i]$;
        $f\_sum[k] \leftarrow f\_sum[k] + temp[k][i]$;
    **end for**
**end for**
$f(x[B]) \leftarrow f\_sum[B]$;

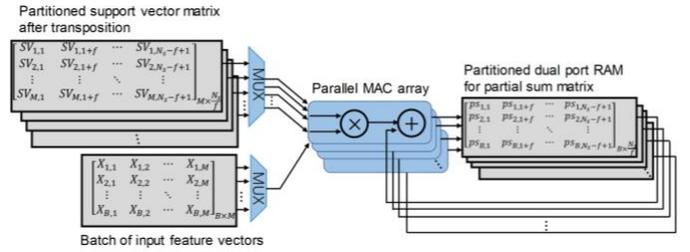

Fig. 8. Hardware structure for calculating the partial sum matrix with batch processing.

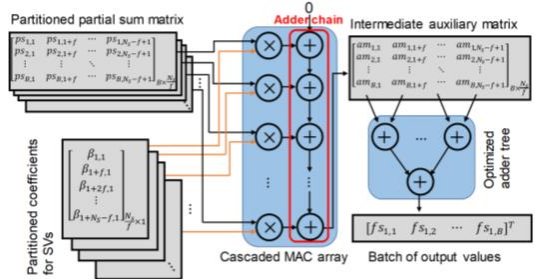

Fig. 9. Hardware structure for calculating the final outputs with batch processing.

The hardware structure of loop L4 and L5 is presented in Fig. 9. Different from the structure in Fig. 8, the massive MAC units are reconstructed to a cascaded MAC array. In every clock cycle, $f$ elements from partial sum matrix and coefficients vector are fetched to the MAC array, while only one output is generated to the intermediate auxiliary matrix at a time. The long adder chain inside the MAC array is heavily pipelined to ensure the initiation interval of 1 clock cycle. It takes $BN_s/f$ cycles to feed all the inputs to the MAC array, however, the latency of the adder chain is not negligible since it is directly proportional to the unroll factor $f$. After the intermediate auxiliary matrix is completely updated, an optimized adder tree will generate the final outputs in serial, the time consumption of this adder tree is trivial since $N_s/f$ is normally very small for large unroll factors.

To verify the effectiveness of batch processing, we apply different batch sizes on Algorithm 3. The unroll factor of 284 is chosen to maximize the use of the available DSP resources on ZCU104. The latency and speedup versus batch size is depicted in Fig. 10(a). We can see that the latency decreases rapidly along with the increase of batch size, and finally converges to about 900 clock cycles. Meanwhile, the speedup increases along with the batch size, and the maximum speedup achieved is 275x with batch size of 40. The hardware utilization is shown in Fig. 10(b). We can see that the DSP, BRAM and FF usage does not change much when the batch size increases. Only the LUT consumption slightly increases since the storage requirement for intermediate values like partial sum matrix and intermediate auxiliary matrix is proportional to batch size. The overall hardware utilization for large batch size does not impose heavy burden to the resources, which proves our proposed batch processing method is also area efficient for hardware implementation.

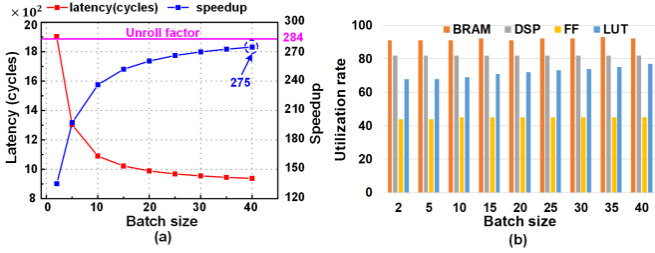

Fig. 10. (a) Speedup and latency versus batch size, (b) hardware utilization rate on ZCU104 versus batch size.

### D. Implementation Results on ZC706 and ZCU104

Next, we implement linear SVR decision function on two different FPGA platforms based on the proposed optimization methods. Two FPGA boards are Xilinx ZC706 and ZCU104 as shown in Fig. 11. The post-implementation resource utilization is shown in Table I, it can be observed that the resources are used adequately for both platforms. The performances of two FPGA boards are shown in Table II, which also includes a software implementation based on widely used LIBSVM running on a computer with i7-5960x CPU and 32 GB RAM. From Table II, we can see that the software implementation with LIBSVM needs 19.41 seconds for the post-processing of 96,100 BGSs from 38.44-km FUT when it works at 3GHz, taking up 18~87.8% of total measurement time. On the contrast, our implementation with ZC706 can complete the post-processing in 1.98 second, while the power consumption of the FPGA development board is only 14.43W when it works at 100MHz, taking up 2.2~42.3% of measurement time. Furthermore, the implementation with ZCU104 completes the post-processing in 0.46 seconds when it works at 200MHz, taking up 0.52~14.5% of measurement time. The power consumption is 26.5W. The working frequency difference between ZC706 and ZCU104 is due to the different manufacturing technology by the two FPGAs, and advanced technology can enable higher working frequency. The equivalent performance of the three platforms are 2.48GFLOPS, 24.3GFLOPS and 104GFLOPS, respectively. The results prove that the hardware accelerators can achieve real-time post-processing for the BOTDA data, which are 9.8x and 42x faster than the software implementation. Meanwhile, two implementations also achieve 95.1x and 226.1x energy efficiency compared with i7-5960x, which could save plenty of energy in all-day monitoring environments.

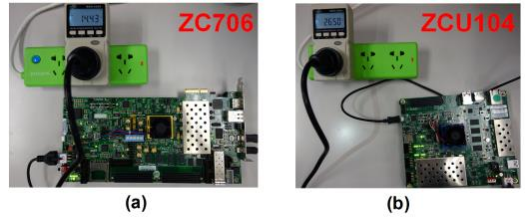

Fig. 11. FPGA boards of (a) Xilinx ZC706, (b) ZCU104.

TABLE I POST-IMPLEMENTATION RESOURCE UTILIZATION OF ZC706 AND ZCU104

|  | Xilinx ZC706 | | | Xilinx ZCU104 | | |
| --- | --- | --- | --- | --- | --- | --- |
|  | Used | Available | Utilization rate | Used | Available | Utilization rate |
| BRAM | 290.5 | 545 | 53.30 | 286 | 312 | 91.67 |
| DSP | 710 | 900 | 78.89 | 1421 | 1728 | 82.23 |
| LUT | 111415 | 218600 | 50.97 | 149623 | 230400 | 64.94 |
| FF | 73213 | 437200 | 16.75 | 199529 | 460800 | 43.30 |

TABLE II PERFORMANCE COMPARISON BETWEEN SOFTWARE IMPLEMENTATION AND TWO FPGA PLATFORMS

| Platform | Intel i7-5960x | Xilinx ZC706 | Xilinx ZCU104 |
| --- | --- | --- | --- |
| Technology | 22nm | 28nm | 16nm |
| Frequency | 3.0 GHz | 100 MHz | 200 MHz |
| Power | 140 W | 14.43 W | 26.50 W |
| Latency(sec) | 19.41 | 1.98 | 0.46 |
| $T_{nn}/T$ | 18~87.8% | 2.2~42.3% | 0.52~14.5% |
| Performance (GFLOPS) | 2.48 | 24.3 | 104 |
| Energy efficiency | 1x | 95.1x | 221.6x |

### E. Theoretical Analysis and Discussions

In Part B and C, we have systematically optimized the original linear SVR decision function for hardware implementation. Loop distribution and loop interchange enable efficient pipeline strategy to be used for partial sum calculation, loop unroll further greatly reduces the latency through parallelizing the MAC operations. Furthermore, the batch processing method makes the latency of the long adder chain shared by multiple inputs, which makes the linear scaling of speedup holds approximately. These optimization techniques make the SVR decision function very suitable to be mapped to FPGA, which are also reflected in the hardware structures in Fig. 8 and Fig. 9. If we further analyze Algorithm 3, we find that we have actually transformed the partial sum matrix calculation and final sum vector calculation to matrix-matrix multiplication and matrix-vector multiplication as follows:

$$\begin{bmatrix} ps_{1,1} & ps_{1,2} & ps_{1,3} & ps_{1,4} & ps_{1,5} & ps_{1,6} & \cdots & ps_{1,N_S} \\ ps_{2,1} & ps_{2,2} & ps_{2,3} & ps_{2,4} & ps_{2,5} & ps_{2,6} & & ps_{2,N_S} \\ \vdots & & & & & & \ddots & \vdots \\ ps_{B,1} & ps_{B,2} & ps_{B,3} & ps_{B,4} & ps_{B,5} & ps_{B,6} & & ps_{B,N_S} \end{bmatrix}$$

$$= \begin{bmatrix} X_{1,1} & X_{1,2} & \cdots & X_{1,M} \\ X_{2,1} & X_{2,2} & \cdots & X_{2,M} \\ \vdots & \vdots & & \vdots \\ X_{B,1} & X_{B,2} & \cdots & X_{B,M} \end{bmatrix} \begin{bmatrix} SV_{1,1} & SV_{1,2} & SV_{1,3} & SV_{1,4} & SV_{1,5} & SV_{1,6} & \cdots & SV_{1,N_S} \\ SV_{2,1} & SV_{2,2} & SV_{2,3} & SV_{2,4} & SV_{2,5} & SV_{2,6} & & SV_{2,N_S} \\ \vdots & & & & & & \ddots & \vdots \\ SV_{M,1} & SV_{M,2} & SV_{M,3} & SV_{M,4} & SV_{M,5} & SV_{M,6} & & SV_{M,N_S} \end{bmatrix} \quad (4.8)$$

$$\begin{bmatrix} fs_{1,1} \\ fs_{2,1} \\ \vdots \\ fs_{B,1} \end{bmatrix} = \begin{bmatrix} ps_{1,1} & ps_{1,2} & ps_{1,3} & ps_{1,4} & ps_{1,5} & ps_{1,6} & \cdots & ps_{1,N_S} \\ ps_{2,1} & ps_{2,2} & ps_{2,3} & ps_{2,4} & ps_{2,5} & ps_{2,6} & \cdots & ps_{2,N_S} \\ \vdots & & & & & & \ddots & \vdots \\ ps_{B,1} & ps_{B,2} & ps_{B,3} & ps_{B,4} & ps_{B,5} & ps_{B,6} & \cdots & ps_{B,N_S} \end{bmatrix} \begin{bmatrix} \beta_{1,1} \\ \beta_{2,1} \\ \beta_{3,1} \\ \beta_{4,1} \\ \beta_{5,1} \\ \beta_{6,1} \\ \vdots \\ \beta_{N_S,1} \end{bmatrix} + b \quad (4.9)$$

For matrix-matrix multiplication in Equation (4.8), we tile the support vector matrix into small blocks and the input vectors multiply each block in serial. The partial sum matrix is also tiled accordingly. For the matrix-vector multiplication in Equation (4.9), the coefficients vector for support vectors also needs to be partitioned to maintain same level of parallelism. As a result, the two operations are both heavily parallelized, which could take the advantage of massive DSP resources and dual port RAMs on FPGA. To be more specific, the parallel MAC array for matrix-matrix multiplication and cascaded MAC array for matrix-vector multiplication are based on same amount of DSP resources, making our implementation achieve very high hardware utilization efficiency since almost no DSP resources are idle during the computation.

## V. Conclusion

In this paper, a new temperature prediction method for BOTDA system based on SVR is proposed. Unlike SVC which can only predict discrete temperatures, SVR can output continuous values from the measured BOTDA data. We experimentally verify that SVR can achieve comparable performance as SVC under different SNRs. From the hardware perspective, SVR is more hardware friendly than SVC. To accelerate the processing speed of SVR, linear SVR decision function is optimized systematically. The loop-carried dependence in the loop iterations is eliminated by loop distribution and loop interchange. Therefore, the pipeline efficiency of the nested loop is improved. We also propose a batch processing method to further decrease the latency. Using the proposed optimization methods, linear SVR decision function is implemented on two FPGA boards Xilinx ZC706 and ZCU104 to process 96,100 BGSs from 38.44-km FUT acquired from a BOTDA system. Our hardware accelerator can achieve up to 42x speedup compared with the software implementation on an i7-5960x computer. The post-processing time for 96,100 BGSs along 38.44-km FUT is only 0.46 seconds with ZCU104, which makes our implementation capable of real-time prediction. Meanwhile, the power consumption of FPGA is also much lower than a high-end CPU, making the energy efficiency of our FPGA implementation up to 226.1x higher than the software implementation based on LIBSVM.